\documentclass[12pt]{iopart}
\usepackage{iopams,epsfig,color}

\newcommand{\POLAL}[1]{#1}

\newcommand{\APJ}{{\it Astrophys.~J.\ }}
\newcommand{\AAP}{{\it Astron.\ Astrophys.\ }}
\newcommand{\MNRAS}{{\it Mon.\ Not.\ R.~Astron.\ Soc.\ }}
\newcommand{\astroph}[1]{{\it Preprint} astro-ph/#1} 
\newcommand{\PRE}{{\it Phys.\ Rev.~E\ }}
\newcommand{\PF}{{\it Phys.\ Fluids\ }}
\newcommand{\PP}{{\it Phys.\ Plasmas\ }}
\newcommand{\JGR}{{\it J.~Geophys.~Res.\ }}
\newcommand{\JFM}{{\it J.~Fluid Mech.\ }}

\newcommand{\exref}[1]{\eref{#1}}
\newcommand{\exsdash}[2]{\eref{#1}-\eref{#2}}

\newcommand{\eqref}[1]{\eref{#1}}

\newcommand{\eqsdash}[2]{\eref{#1}-\eref{#2}}
\newcommand{\eqsand}[2]{\eref{#1} and \eref{#2}}
\newcommand{\Eqref}[1]{Equation \eref{#1}}

\newcommand{\Eqsdash}[2]{Equations \eref{#1}-\eref{#2}}

\newcommand{\figref}[1]{figure \ref{#1}}

\newcommand{\secref}[1]{\S\ref{#1}}
\newcommand{\secsref}[2]{\S\S\ref{#1},\ref{#2}}
\newcommand{\apref}[1]{\ref{#1}} 

\newcommand{\bea}{\begin{eqnarray}}
\newcommand{\eea}{\end{eqnarray}}
\newcommand{\lt}{\left}
\newcommand{\rt}{\right}
\newcommand{\bl}{\bigl}

\newcommand{\dd}{\partial}
\newcommand{\vdel}{\bnabla}
\newcommand{\dpar}{{\dd\over\dd z}}

\newcommand{\vr}{\bi{r}}

\newcommand{\vu}{\bi{u}}
\newcommand{\vE}{\bi{E}}
\newcommand{\dvE}{\delta\vE}

\newcommand{\vj}{\bi{j}}
\newcommand{\vB}{\bi{B}}
\newcommand{\vb}{\hat{\bi{b}}}
\newcommand{\vz}{\hat{\bi{z}}}

\newcommand{\dB}{\delta B}
\newcommand{\dE}{\delta E}

\newcommand{\dvB}{\delta\vB}

\newcommand{\drho}{\delta\rho}

\newcommand{\Epar}{E_\parallel}

\newcommand{\vuperp}{\vu_\perp}
\newcommand{\vue}{\vuperp}
\newcommand{\uperp}{u_\perp}
\newcommand{\upar}{u_\parallel}

\newcommand{\dvBperp}{\dvB_\perp}
\newcommand{\dBperp}{\delta B_\perp}
\newcommand{\dBpar}{\delta B_\parallel}
\newcommand{\Dpar}{\vb\cdot\vdel}
\newcommand{\vdperp}{\vdel_\perp}
\newcommand{\dperp}{{\nabla}_{\perp}}

\newcommand{\const}{\mathrm{const}}

\newcommand{\kpar}{k_\parallel}
\newcommand{\kparo}{k_{\parallel 0}}
\newcommand{\omegao}{\omega_0}
\newcommand{\kparA}{k_{\parallel A}}
\newcommand{\kperp}{k_\perp}
\newcommand{\mfp}{\lambda_\mathrm{mfp}}

\newcommand{\ul}{\delta u_{\perp\lambda}}
\newcommand{\dBl}{\delta B_{\lambda}}
\newcommand{\dBpl}{\delta B_{\perp\lambda}}
\newcommand{\dBparl}{\delta B_{\parallel\lambda}}
\newcommand{\dnl}{\delta n_{\lambda}}
\newcommand{\epsn}{\varepsilon_{n}}
\newcommand{\epsB}{\varepsilon_{B}}
\newcommand{\xip}{\xi_\perp}
\newcommand{\xil}{\xi_{\perp\lambda}}
\newcommand{\thl}{\theta_\lambda}
\newcommand{\taul}{\tau_\lambda}
\newcommand{\lpar}{l_{\parallel}}
\newcommand{\lparl}{l_{\parallel\lambda}}

\newcommand{\urms}{u_\mathrm{rms}}
\newcommand{\lf}{L}
\newcommand{\dBrms}{\delta B_\mathrm{rms}}
\newcommand{\vths}{v_{\mathrm{th}s}}
\newcommand{\vthi}{v_{\mathrm{th}i}}

\newcommand{\vv}{\bi{v}}
\newcommand{\vperp}{v_\perp}
\newcommand{\vpar}{v_\parallel}
\newcommand{\nui}{\nu_{ii}}
\newcommand{\nupar}{\nu_\parallel}
\newcommand{\kappar}{\kappa_\parallel}

\newcommand{\pperp}{p_\perp}

\newcommand{\ppar}{p_{\parallel}}

\newcommand{\dpperp}{\delta\pperp}
\newcommand{\dpperpi}{\delta p_{\perp i}}
\newcommand{\dpperpe}{\delta p_{\perp e}}
\newcommand{\dppar}{\delta\ppar}

\newcommand{\Tperp}{T_{\perp0}}
\newcommand{\Tpar}{T_{\parallel0}}

\newcommand{\fMs}{F_{0s}}
\newcommand{\fMi}{F_{0i}}
\newcommand{\fMe}{F_{0e}}
\newcommand{\dfs}{\delta f_s}
\newcommand{\dfi}{\delta f_i}
\newcommand{\dfe}{\delta f_e}

\renewcommand{\ne}{n_{0}}
\renewcommand{\ni}{n_{0}}
\newcommand{\Ti}{T_{0}}

\newcommand{\dni}{\delta n_i}
\newcommand{\dne}{\delta n_e}
\newcommand{\dTi}{\delta T_i}
\newcommand{\dTe}{\delta T_e}

\begin{document}


\title[Interplanetary and interstellar plasma turbulence]{Interplanetary and interstellar 
plasma turbulence}

\author{A A Schekochihin$^{1,2}$, S C Cowley$^{2,3}$ and W Dorland$^4$}

\address{$^1$ DAMTP, University of Cambridge, 
Cambridge~CB3~0WA, UK}
\address{$^2$ Department of Physics, Imperial College London, 
London~SW7~2BW, UK}
\address{$^3$ Department of Physics and Astronomy, 
UCLA, Los Angeles, CA~90095-1547, USA}
\address{$^4$ Department of Physics, 
University of Maryland, College Park, MD~20742-3511, USA}

\ead{as629@damtp.cam.ac.uk}

\begin{abstract}

Theoretical approaches to low-frequency magnetized turbulence 
in collisionless and weakly collisional astrophysical plasmas are 
reviewed. The proper starting point for an analytical description of these 
plasmas is kinetic theory, not fluid equations. 
The anisotropy of the turbulence is used to systematically derive 
a series of reduced analytical models. 
Above the ion gyroscale, it is shown rigourously 
that the Alfv\'en waves decouple from 
the electron-density and magnetic-field-strength fluctuations and satisfy the 
Reduced MHD equations. 
The density and field-strength fluctuations (slow waves and 
the entropy mode in the fluid limit), determined kinetically, 
are passively mixed by the Alfv\'en waves. 
The resulting hybrid fluid-kinetic description 
of the low-frequency turbulence is valid independently of collisionality. 
Below the ion gyroscale, the turbulent cascade is partially 
converted into a cascade of kinetic Alfv\'en waves, 
damped at the electron gyroscale. This cascade is 
described by a pair of fluid-like equations, which are a reduced 
version of the Electron MHD. 
The development of these theoretical models is motivated by 
observations of the turbulence in the solar wind and interstellar 
medium. In the latter case, the turbulence is spatially inhomogeneous 
and the anisotropic Alfv\'enic turbulence in the presence of a strong 
mean field may coexist with isotropic MHD turbulence that has no mean field. 

\end{abstract}

\submitto{\PPCF {\rm on 31 May 2006, now accepted and scheduled to be published in May 2007}}

\section{Introduction}

Rapid progress in astronomical instrumentation 
has made it possible to observe astrophysical plasmas with 
ever greater spatial resolution. This has allowed astronomers 
to probe not only the bulk, {\em large-scale} motions and fields 
but also to measure, either directly or via line-of-sight integrated 
quantities associated with the emission and propagation of light, 
the {\em small-scale} fluctuations of plasma velocity, density, 
magnetic and electric fields. 
These {\em turbulent} fluctuations existing in 
a broad range of scales are 
a common property of astrophysical plasmas. 
While astrophysical turbulence occurs in a 
variety of vastly differing conditions, its physical characterization 
is based on a number of universal features. 
In most cases, the source of energy is 
in the form of random stirring or instabilities associated 
with the scale of the astrophysical object of interest. 
The energy injected at 
large scales cascades to much (typically many orders of magnitude) smaller 
scales to be dissipated into heat. A signature property 
of the turbulent cascade that connects these vastly disparate scales 
is power-law spectra of the fluctuating quantities. These have been 
observed in the solar wind (SW), e.g.~\cite{Leamon_etal,Bale_etal,Horbury_etal_review}, 
the interstellar medium (ISM) \cite{Armstrong_Rickett_Spangler,Minter_Spangler,Lazio_etal_review}, 
galaxy clusters \cite{Schuecker_etal,Vogt_Ensslin2}, etc. 
In all of the cited examples, the reported spectra had, or were 
consistent with, Kolmogorov scaling $k^{-5/3}$. 

In this paper, we shall concentrate on the SW and ISM and 
outline both the qualitative understanding that 
currently exists of the turbulence in these media and 
a formal mathematical description of this turbulence 
that must underlie the future analytical and numerical investigations 
of it. The proper starting point for such a description is the kinetic 
plasma theory because the turbulent plasmas we are interested in 
are either collisionless (in the SW, the particle mean free path 
is comparable to the distance from the Sun to the Earth) 
or only weakly collisional, 
meaning that the mean free path $\mfp$ exceeds the ion gyroradius $\rho_i$ 
(in the ISM, $\mfp\sim10^{12}$~cm, $\rho_i\sim10^{9}$~cm). 

In many cases, it is plausible to think of plasma 
turbulence at scales much smaller than the energy-injection scale as an ensemble 
of interacting MHD waves propagating along a dynamically strong 
background magnetic field (the mean field) associated with the 
large scales \cite{Kraichnan}. Goldreich and Sridhar \cite{GS95} 
(henceforth, GS) conjectured that in such a turbulence, (i) all electromagnetic 
perturbations are strongly anisotropic, 
so that the characteristic wavenumbers along the field 
are much smaller than those across it, $\kpar\ll\kperp$; 
and (ii) the interactions between the Alfv\'en waves are 
strong, i.e., the Alfv\'en time and the 
nonlinear interaction time are comparable to each other: 
\bea
\label{crit_bal}
\omega\sim\kpar v_A\sim\kperp\uperp,
\eea
where $\omega$ is the typical frequency of perturbations, 
$v_A$ is the Alfv\'en speed, and 
$\uperp$ is the velocity fluctuation perpendicular to the 
mean field. This assumption, known as {\em the critical balance}, 
removed dimensional ambiguity from the MHD turbulence theory and 
led to the Kolmogorov scaling of the 
Alfv\'en-wave energy spectrum, $\kperp^{-5/3}$ 
and to the relation $\kpar\sim\kperp^{2/3}$ 
(for a historical review, see \cite{SC_mhdbook}; 
in \apref{ap_GS}, we give a brief outline of the GS theory 
and related scaling arguments for MHD turbulence). 

The anisotropy of MHD turbulence is supported by observations 
of the SW 
the ISM 
(see reviews \cite{Horbury_etal_review,Lazio_etal_review}) 
and by numerical simulations \cite{Maron_Goldreich,CLV_aniso,Mueller_Biskamp_Grappin,Cho_Lazarian_EMHD}.
In what follows, this anisotropy emerges as the key simplifying 
feature used to derive a reduced 
version of the plasma kinetic theory that describes low-frequency 
MHD turbulence. 
This is done in \secref{sec_SW}, where our exposition is motivated 
by the observations of the collisionless SW. We show how 
the descriptions known as Reduced MHD, Kinetic MHD, Electron MHD 
and Gyrokinetics fit into a single theoretical framework. 
In \secref{sec_ISM}, we explain how the same approach works 
for the turbulence in parts of the ISM and how this type of turbulence 
differs from the isotropic MHD turbulence, which does not have 
a mean field. We argue that the latter kind of turbulence may also 
be present in the ISM and in galaxy clusters. 

\section{Solar wind and the collisionless MHD turbulence}
\label{sec_SW}

\begin{figure}[t]
\centerline{\epsfig{file=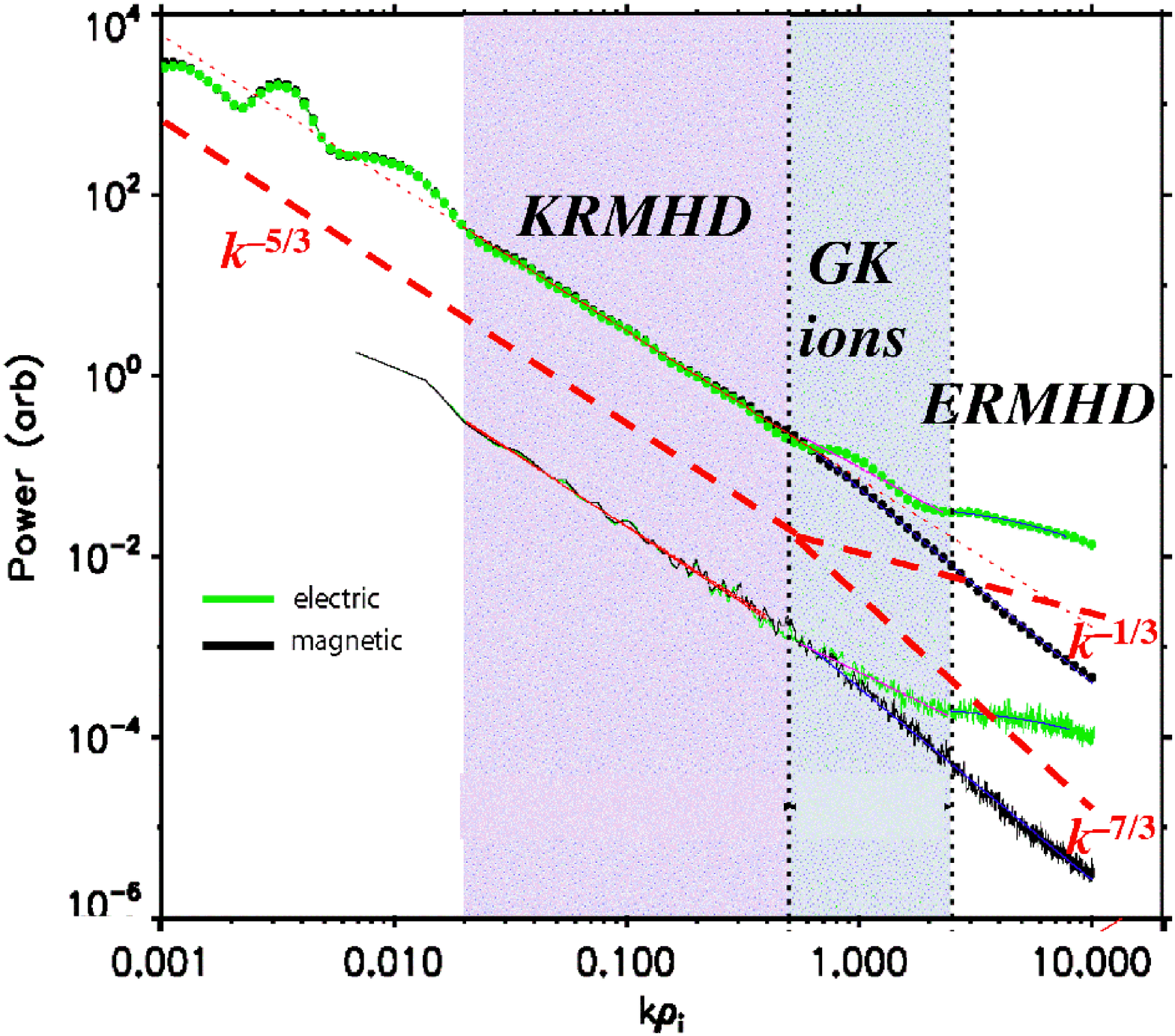,width=3.5in}}
\caption{\label{fig_bale} Spectra of electric and magnetic fluctuations in the 
SW --- adapted with permission from figure 3 of \cite{Bale_etal} (copyright 2005 by the American 
Physical Society). The bold dashed (red) lines are reference slopes added by us. 
We also inserted ``KMHD'', ``GK ions'' and ``ERMHD''  
to indicate the scale intervals where these analytical descriptions are 
valid: $\kperp\ll\rho_i^{-1}$ (see \secref{sec_AW}), 
$\kperp\sim\rho_i^{-1}$ (see \secref{sec_GK}) and 
$\rho_i^{-1}\ll\kperp\ll\rho_e^{-1}$ (see \secref{sec_ERMHD}), respectively.}
\end{figure}

Spectra of electromagnetic fluctuations 
in the SW extend across a broad range of collisionless 
scales. Above the ion gyroscale ($\rho_i\sim100$~km), 
the spectra of the electric and magnetic field measured 
by spacecraft at 1 AU from the Sun 
fit the $k^{-5/3}$ law and follow each other 
with remarkable precision (\figref{fig_bale}). 
Since the electric field is directly related 
to the plasma velocity (at scales above $\rho_i$, 
it is the $\vE\times\vB$ drift velocity), this 
can be interpreted as a signature of Alfv\'enic turbulence. 
How do we describe such a turbulence at collisionless scales? 
Let us first consider scales larger than the ion gyroscale. 

Note that the plasma beta $\beta_i$ is taken 
to be order unity in what follows, as is appropriate both for 
the SW and the ISM. It is useful to remember that in this regime, 
the ion inertial scale is comparable to $\rho_i$. 

\subsection{Kinetic MHD}
\label{sec_KMHD}

For $k\rho_i\ll1$, the magnetic field 
impedes free particle motion across the field lines and the kinetic theory 
reduces to the so-called {\em Kinetic MHD (KMHD)} \cite{Kulsrud_HPP}, 
which has most features of the MHD description, but 
allows for anisotropic pressure:  
\bea
\label{KMHD_rho}
{\rmd\rho\over\rmd t} = -\rho\vdel\cdot\vu,\\
\label{KMHD_u}
\rho\,{\rmd\vu\over\rmd t} = -\vdel\lt(\pperp+{B^2\over8\pi}\rt) 
+ \vdel\cdot\lt[\vb\vb\lt(\pperp-\ppar\rt)\rt] + {\vB\cdot\vdel\vB\over4\pi},\\
\label{KMHD_B}
{\rmd\vB\over\rmd t} = \vB\cdot\vdel\vu - \vB\vdel\cdot\vu,
\eea
where $\rho$ is mass density, $\vu$ velocity, $\vB$ magnetic field, 
$\vb=\vB/B$, and $\rmd/\rmd t=\dd/\dd t + \vu\cdot\vdel$. 
The pressure tensor is calculated kinetically: 
$\pperp=\sum_s m_s\int\rmd^3\vv(\vperp^2/2)f_s$ 
and $\ppar=\sum_s m_s\int\rmd^3\vv\bl(\vpar-\upar)^2f_s$, 
where the distribution function $f_s(t,\vr,\vperp,\vpar)$ 
satisfies 
\bea
{Df_s\over Dt} + {1\over B}{DB\over Dt}{\vperp\over2}{\dd f_s\over\dd\vperp}-
\Biggl(\vb\cdot{D\vue\over Dt} + {\vperp^2\over2}{\vb\cdot\vdel B\over B} 
- {q_s\Epar\over m_s}\Biggr){\dd f_s\over\dd\vpar} &=& 0,
\label{KMHD_f}
\eea
where $D/Dt = \dd/\dd t + (\vue+\vpar\vb)\cdot\vdel$ 
and $\vue=\vu-\vu\cdot\vb\vb$. 
In the above, $m_s$ and $q_s=\pm e$ are the mass and charge 
of the particles of species $s$ (ions, electrons). 
The parallel electric field $\Epar$ is determined from the quasineutrality 
condition $\sum_s q_s n_s = 0$, where $n_s=\int\rmd^3\vv\,f_s$ (the number density). 
Note that $\rho=m_i n_i$ and $\upar = (1/n_i)\int\rmd^3\vv\,\vpar f_i$, so \eqref{KMHD_rho} 
and the parallel component of \eqref{KMHD_u} can be derived from \eqref{KMHD_f}. 

We consider a uniform static equilibrium with a straight mean field 
in the $z$ direction, so $\vB = B_0\vz + \dvB$ and  
$\rho = \rho_0 + \drho$, $\pperp = p_0 + \dpperp$, $\ppar = p_0 + \dppar$, 
where $B_0$, $\rho_0$, $p_0$ are constant in space and time. 

\subsection{The ordering}
\label{sec_ordering}

The anisotropy of the turbulence allows us to 
systematically expand \eqsdash{KMHD_rho}{KMHD_f} in $\epsilon\sim\kpar/\kperp$. 
The key step in setting up such an expansion is to estimate 
the strength of the fluctuations by adopting the critical-balance 
conjecture \exref{crit_bal}, 
but as an ordering assumption rather than a detailed scaling prescription: 
this means that the wave propagation terms are assumed to be same order 
as the nonlinear interaction terms (the turbulence is strong). 
This leads to the following ordering:
\bea
{\drho\over\rho_0}
\sim {\uperp\over v_A} \sim {\upar\over v_A}
\sim {\dpperp\over p_0} \sim {\dppar\over p_0} 
\sim {\dBperp\over B_0} \sim {\dBpar\over B_0} 
\sim \epsilon, 
\label{RMHD_ordering}
\eea
where $v_A=B_0/\sqrt{4\pi\rho_0}$. 
Two auxiliary ordering assumptions have been made: 
(i) the perpendicular velocity and magnetic-field
fluctuations are Alfv\'enic ($\dBperp/B_0\sim\uperp/v_A$); 
(ii) the Alfv\'enic fluctuations 
are same order as magnetic-field-strength ($\dBpar$),  
density and pressure fluctuations 
(in the collisional MHD limit, these correspond to the slow waves 
and the entropy mode). 
The validity of the latter assumption 
depends on how the turbulence is stirred. 
In astrophysical contexts, the large-scale energy input 
may be assumed to inject comparable power 
into all types of fluctuations. 
We also assume that the characteristic frequency of the 
fluctuations is $\omega\sim\kpar v_A$.
The fast waves are, thus, ordered out, because 
their frequency is $\sim\kpar v_A/\epsilon$. 

\subsection{Alfv\'en-wave turbulence: Reduced MHD}
\label{sec_AW}

The Alfv\'enic fluctuations are 
two-dimensionally solenoidal: since 
$\vdel\cdot\vu=O(\epsilon^2)$ [from \eqref{KMHD_rho}] and $\vdel\cdot\dvB=0$ exactly, 
separating the $O(\epsilon)$ part of these divergences gives 
$\vdperp\cdot\vuperp=\vdperp\cdot\dvBperp=0$. 
Therefore, to lowest order, we may write 
$\vuperp = \vz\times\vdperp\Phi$ and 
${\dvBperp/\sqrt{4\pi\rho_0}} = \vz\times\vdperp\Psi$. 
Equations for the scalar fields 
$\Phi$ and $\Psi$ (the stream and flux functions) 
are obtained by substituting these expressions into 
the perpendicular parts of \eqsand{KMHD_u}{KMHD_B} 
--- of the former the curl is taken to 
annihilate the pressure term. Keeping only the terms of the lowest order, 
$O(\epsilon^2)$, we get 
\bea
\label{RMHD_Phi}
{\dd\over\dd t}\dperp^2\Phi + \lt\{\Phi,\dperp^2\Phi\rt\} 
= v_A\dpar\dperp^2\Psi + \lt\{\Psi,\dperp^2\Psi\rt\},\quad\\
\label{RMHD_Psi}
{\dd\over\dd t}\,\Psi + \lt\{\Phi,\Psi\rt\} = v_A\dpar\Phi,
\eea
where $\lt\{\Phi,\Psi\rt\}=\vz\cdot(\vdperp\Phi\times\vdperp\Psi)$ 
and we have taken into account that, to lowest order,  
\bea
\label{dd_def}
\fl
{\rmd\over\rmd t} = {\dd\over\dd t} + \vuperp\cdot\vdperp={\dd\over\dd t} + \lt\{\Phi,\cdots\rt\},
\ 
\Dpar = \dpar + {\dvBperp\over B_0}\cdot\vdperp 
= \dpar + {1\over v_A}\lt\{\Psi,\cdots\rt\}.
\eea
The closed system \eqsdash{RMHD_Phi}{RMHD_Psi}, 
known as the {\em Reduced MHD (RMHD)}, was
derived originally from the fluid MHD equations for the studies 
of stability of fusion plasmas \cite{Strauss,Kadomtsev_Pogutse}. 
We have now shown that {\em the Alfv\'en-wave cascade in a collisionless plasma 
is described by the RMHD equations \exsdash{RMHD_Phi}{RMHD_Psi} all the way 
down to the ion gyroscale.}
The Alfv\'en waves are decoupled from the other fluctuation modes: 
density and magnetic-field-strength fluctuations (slow waves and 
entropy fluctuations in the collisional limit; cf.~\cite{GS97,Lithwick_Goldreich}). 

Introducing {\em Elsasser fields} $\zeta^\pm=\Phi\pm\Psi$, we may 
rewrite \eqsdash{RMHD_Phi}{RMHD_Psi} as follows
\bea
\fl
{\dd\over\dd t}\dperp^2\zeta^\pm \mp 
v_A\dpar\dperp^2\zeta^\pm = 
-{1\over2}\lt[\lt\{\zeta^+,\dperp^2\zeta^-\rt\}
+ \lt\{\zeta^-,\dperp^2\zeta^+\rt\}
\mp\dperp^2\lt\{\zeta^+,\zeta^-\rt\}\rt].
\label{eq_zeta}
\eea
Thus, the RMHD, like the MHD, supports 
wave packets of arbitrary shape and magnitude 
propagating in one direction at the Alfv\'en speed $v_A$: 
if $\zeta^-=0$ or $\zeta^+=0$, 
the nonlinear terms vanish and the exact solution for the 
other Elsasser potential is 
$\zeta^\pm = f^\pm(x,y,z\mp v_A t)$, 
where $f^\pm$ is an arbitrary function. 
The Alfv\'en-wave cascade is a result of interactions between  
counterpropagating wave packets \cite{Kraichnan}. 

It is this Alfv\'enic component of the plasma turbulence to which 
the GS scaling theory of MHD turbulence (\apref{ap_GS}) applies.  
In the SW, it is observed via {\em in situ} 
measurements of the fluctuating magnetic and electric fields \cite{Bale_etal} 
(see \figref{fig_bale}). 
The latter directly probe the velocity fluctuations because, 
to lowest order in~$\epsilon$, 
$\vuperp = c\vE\times\vB/B^2 = (c/B_0)\vz\times\vdperp\phi$, 
where $\phi$ is the scalar potential. Clearly, $\Phi=c\phi/B_0$. 

\subsection{Density and magnetic-field-strength fluctuations: Kinetic Reduced MHD}
\label{sec_nB}

In order to determine $\dne$ and $\dBpar$,
we must use the kinetic equation~\exref{KMHD_f}.\footnote{These 
quantities cannot be derived from \eqref{KMHD_rho} and 
the parallel part of \eqref{KMHD_B} because 
(i) \eqref{KMHD_rho} has already been used to determine $\vdel\cdot\vu$, 
a $O(\epsilon^2)$ quantity; 
(ii) the parallel part of \eqref{KMHD_B} contains $\upar$, 
whose evolution equation, 
the parallel part of \eqref{KMHD_u}, requires $\dpperp-\dppar$, 
so $\upar$ can only be calculated kinetically.} 
The lowest-order (equilibrium) distribution is taken 
to be a Maxwellian: $\fMs = n_0\,\rme^{-v^2/\vths^2}/(\pi\vths^2)^{3/2}$, 
where $\vths=(2T_0/m_s)^{1/2}$ is the thermal speed of species $s$ 
and $T_0$ is temperature.\footnote{The assumption of an isotropic 
equilibrium was implicit when we adopted an isotropic zeroth-order pressure $p_0$ 
at the end of \secref{sec_KMHD}. Strictly speaking, in a collisionless plasma
such as the SW, the equilibrium distribution does not have to be 
Maxwellian or isotropic. The conservation of the first adiabatic 
invariant, $\mu=\vperp^2/2B$, suggests that temperature anisotropy 
with respect to the magnetic-field direction ($\Tperp\neq\Tpar$) may 
exist. Such anisotropy gives rise to several high-frequency 
plasma instabilities \cite{Gary_book} and it is plausible to assume that 
fluctuations associated with them will scatter particles 
and limit the anisotropy (e.g., \cite{Kellogg}). 
While there is no definitive analytical theory quantifying this idea, 
it has some support in the SW observations that indicate that the 
core particle distribution is only moderately anisotropic \cite{Marsch_Ao_Tu}. 
We believe, therefore, that assuming a Maxwellian equilibrium 
is an acceptable simplification. We also take $T_{0i}=T_{0e}$ 
(generalising to $T_{0i}\neq T_{0e}$ is straightforward). 
Note that in plasmas such as the ISM, where collisions are weak but 
non-negligible (\secref{sec_weak_coll}), 
the Maxwellian equilibrium is rigourously justifiable 
if the ion collision rate is ordered $\nui\sim\omega$ 
within the $\epsilon$ expansion~\cite{Howes_etal}.} 
We let $f_s=\fMs + \dfs$, where $\dfs/\fMs\sim\epsilon$ 
and apply the ordering \exref{RMHD_ordering} to the kinetic equation 
\exref{KMHD_f}. 

The electron kinetic equation can be further simplified by a 
subsidiary expansion in $(m_e/m_i)^{1/2}$ \cite{Snyder_Hammett}. 
To lowest order,  
\bea
\label{eq_dfe}
\vpar\lt(\Dpar\dfe + {e\over T_0}\,\Epar\fMe\rt) = 0.
\eea
Since $\int\rmd^3\vv\,\dfe = \dne$, the inhomogeneous solution of 
this equation is $\dfe=(\dne/\ne)\fMe$ 
(the electrons are isothermal). The homogeneous solution 
satisfies $\Dpar\dfe = 0$, i.e., it is constant along 
the perturbed field lines and is constant 
everywhere if the field lines are assumed to be stochastic. 
Thus, $\Epar=-(T_0/e\ne)\vb\cdot\vdel\dne$. 
Substituting this into the ion kinetic equation, 
we have, to lowest order, $O(\epsilon^2)$, 
\bea
{\rmd\over\rmd t}\lt(\dfi - {\vperp^2\over\vthi^2}{\dBpar\over B_0}\fMi\rt) +
\vpar\Dpar\lt(\dfi + {\dne\over\ne}\fMi\rt) = 0.
\label{eq_dfi}
\eea
Finally, we calculate $\dne$ and $\dBpar$.
From quasineutrality, $\dne=\dni$, so
\bea
\label{KRMHD_n}
{\dne\over\ne} = {1\over\ni}\int\rmd^3\vv\,\dfi.
\eea
To calculate $\dBpar$, we first revisit the the perpendicular part of \eqref{KMHD_u}.  
In the lowest order, $O(\epsilon)$, it 
reduces to the perpendicular pressure balance:
$\vdperp\lt(\dpperp + {B_0\dBpar/4\pi}\rt) = 0$, 
whence $\dBpar = -(4\pi/B_0)\dpperp$ 
(this is why the fast waves disappear under our ordering). 
Now $\dpperp=\dpperpe+\dpperpi$. 
Using $\dfe=(\dne/\ne)\fMe$ to get $\dpperpe=T_0\dne$,  
equation \exref{KRMHD_n} to express $\dne$, and calculating 
$\dpperpi$ from $\dfi$, we find 
\bea
\label{KRMHD_B}
{\dBpar\over B_0} = -{\beta_i\over2}{1\over\ni}\int\rmd^3\vv\lt(1+{\vperp^2\over\vthi^2}\rt)\dfi,
\eea
where $\beta_i=8\pi\ni T_0/B_0^2=\vthi^2/v_A^2$. 
Note that $\upar=(1/\ni)\int\rmd^3\vv\,\vpar\dfi$ is not required to solve the equations, 
but can be calculated from the solution. 

Together with \eqsdash{RMHD_Phi}{RMHD_Psi},  
\eqsdash{eq_dfi}{KRMHD_B} 
form a closed system that describes the anisotropic turbulence 
above the ion gyroscale in a collisionless magnetized plasma.  
We shall refer to this hybrid fluid-kinetic theory as {\em Kinetic RMHD (KRMHD).} 
The nonlinearity enters in \eqref{eq_dfi}
via the derivatives defined in \eqref{dd_def} and is due solely to 
interactions with Alfv\'en waves. 
Thus, {\em the cascades of density and magnetic-field-strength fluctuations 
occur via passive mixing by Alfv\'en waves, with no energy 
exchange} (cf.~\cite{GS97,Lithwick_Goldreich}).

\subsection{Parallel and perpendicular cascades}
\label{sec_cascades}

Let us transform \eqref{eq_dfi} 
to the Lagrangian frame associated with the 
velocity field $\vuperp$ of the Alfv\'en waves: 
$(t,\vr)\to(t,\vr_0)$, where $\vr(t,\vr_0) = \vr_0 + \int_0^t\rmd t'\,\vuperp(t',\vr(t',\vr_0))$. 
In this frame, $\rmd/\rmd t$ [defined in \eqref{dd_def}] becomes $\dd/\dd t$. 
\Eqref{KMHD_B} has the Cauchy solution: 
$\vB(t) = [\rho(t)/\rho(0)]\vB(0)\cdot\vdel_0\vr$, where $\vdel_0=\dd/\dd\vr_0$.  
Then 
$\Dpar = \vb(0)\cdot\lt(\vdel_0\vr\rt)\cdot\vdel = 
\vb(0)\cdot\vdel_0 = {\dd/\dd l_0}$,
where $l_0$ is the arc length along the magnetic field line 
taken at $t=0$ [if $\dvBperp(0)=0$, $l_0=z_0$]. 
Thus, in the Lagrangian frame associated with the Alfv\'en waves, 
\eqref{eq_dfi} is linear. It does not, therefore, 
support a cascade of $\dne$ and $\dBpar$ to smaller 
scales parallel to the perturbed magnetic field, i.e., $\vb\cdot\vdel$ 
of these fluctuations does not change with time. 
In contrast, passive mixing by the Alfv\'en waves does cause a 
perpendicular cascade 
of $\dne$ and $\dBpar$ --- i.e., a cascade in $\kperp$.

Unlike \eqref{eq_dfi}, the RMHD equations 
\exsdash{RMHD_Phi}{RMHD_Psi} in the Lagrangian frame 
do not reduce to a linear form, so the Alfv\'en waves 
should develop small scales both across and along the perturbed 
magnetic field. The scale-by-scale critical balance 
\exref{crit_bal} conjectured by GS leads to the relation 
$\kpar\sim\kperp^{2/3}$ (see \apref{ap_GS}). 

Using the linearity of \eqref{eq_dfi} in the Largangian frame, 
it is straightforward to show that density and field-strength fluctuations 
are damped. The dispersion relation is 
\bea
\label{disp_rel}
{\omegao\over|\kparo|\vthi}Z\lt({\omegao\over|\kparo|\vthi}\rt) 
= -2\lt(1 - {1\over2\beta_i} \pm \sqrt{1 + {1\over4\beta_i^2}}\rt),
\eea 
where $Z$ is the plasma dispersion function and $\omegao$ and 
$\kparo$ are the {\em Lagrangian} frequency and wave number 
($\kparo\sim\Dpar$). When $\beta_i\sim1$, all solutions of \eqref{disp_rel} 
have damping rates ${\rm Im}(\omegao)\sim-|\kparo|\vthi\sim 
-|\kparo|v_A$.\footnote{For $\beta_i\gg1$, 
the weakest-damped solution is $\omegao\simeq-i|\kparo|v_A/\sqrt{\pi\beta_i}$. 
This is the anisotropic limit ($\kpar/\kperp\ll1$) 
of the more general effect known as Barnes, or transit-time, damping \cite{Barnes}. 
Note that we carried out the expansion in small $\kpar/\kperp$ 
before taking the high-$\beta$ limit. A more standard approach in the linear 
theory of plasma waves is to leave $\kpar/\kperp$ arbitrary and 
take the high-$\beta$ limit first \cite{Foote_Kulsrud}.} 
If no parallel cascade of $\dne$ and $\dBpar$ develops, 
the parallel wavenumber $\kparo$ of these fluctuations with a given $\kperp$
does not grow with $\kperp$, so, for large enough $\kperp$, it 
is much smaller than the parallel wave number $\kparA\sim\kperp^{2/3}$ 
of the Alfv\'en waves that have the same $\kperp$. 
This means that the damping rate is small compared to the characteristic 
rate $\kparA v_A$ at which the Alfv\'en waves cause $\dne$ and $\dBpar$ 
to cascade to higher $\kperp$. One is then led to conclude that, 
despite the kinetic damping, $\dne$ and $\dBpar$ 
should have perpendicular cascades extending to the ion gyroscale. 

The validity of this conclusion is not quite as obvious 
as it might appear. 
Lithwick and Goldreich \cite{Lithwick_Goldreich} argued that 
the dissipation of $\dne$ and $\dBpar$ at the ion gyroscale 
would lead these fluctuations to become uncorrelated at the 
same parallel scales as the Alfv\'enic fluctuations by which 
they are mixed, i.e., $\kparo\sim\kparA$. The damping rate 
then becomes comparable to the cascade rate, causing the 
cascades of density and field-strength fluctuations to be 
cut off at $\kpar\mfp\sim1$. In the SW, this would 
mean that no such fluctuations should be detected above 
the ion gyroscale. Observational evidence is at odds with 
this conclusion: the density fluctuations appear to follow 
a $k^{-5/3}$ law at $\kperp\rho_i\ll1$ \cite{Celnikier_etal87}, 
as they should if they are passively mixed and not damped 
(see \apref{ap_GS}). The same is true for 
the fluctuations of the field strength \cite{Bershadskii_Sreeni_Bpar,Hnat_Chapman_Rowlands2}.  
It is not clear why Lithwick and Goldreich's argument fails, 
but it is, perhaps, useful to point out two potential pitfalls: 
(i) in order for the dissipation terms, not present in 
\eqsdash{eq_dfi}{KRMHD_B}, to act, the density and field-strength 
fluctuations should reach the ion gyroscale in the first place; 
(ii) the damping rate of these fluctuations, even if $\kparo\sim\kparA$, 
is never much larger than the cascade rate, so it may be necessary 
to have a quantitative calculation of the interplay between 
the kinetic damping, mixing and the dissipation at $\kperp\rho_i\sim1$ 
in order to determine the efficiency of the cascade. 

\subsection{Gyrokinetics} 
\label{sec_GK}

At $\kperp\rho_i\sim1$, the KMHD description breaks down and 
the Alfv\'enic fluctuations are no longer decoupled 
from the kinetic component of the turbulence. They are mixed with the fluctuations 
of the density and magnetic-field strength and dissipated via the 
collisionless damping discussed in \secref{sec_cascades} 
--- the observed flattening of the density-fluctuation spectrum as $\kperp\rho_i$ 
approaches unity \cite{Celnikier_etal87} is likely to be due to this energy 
exchange with the Alfv\'en waves. The damping 
leads to ion heating, an astrophysically interesting problem in its 
own right, e.g., in the theories of coronal heating \cite{Cranmer_vanBallegooijen} 
and accretion discs \cite{Quataert_Gruzinov}. 
The amount of heating suffered by the ions is a nontrivial issue 
because only part of the turbulent energy is dissipated at $\kperp\rho_i\sim1$. 
The rest is converted into a cascade of kinetic Alfv\'en waves (KAW) 
that extends to the electron gyroscale --- a feature observed in the 
SW \cite{Leamon_etal,Bale_etal}.  
Quantitative theory or numerical modeling of the energy dissipation and 
conversion processes at $\kperp\rho_i\sim1$ 
can only be done in the fully kinetic framework. 
However, the anisotropy of the fluctuations 
leads to a substantial simplification of the full 
plasma kinetic theory. If the ordering based on the assumptions of  
anisotropy and critical balance (\secref{sec_ordering}) is applied, 
the plasma kinetics reduce to {\em gyrokinetics} (GK) --- a low-frequency 
limit well known in fusion science \cite{Brizard_Hahm_review}. 
A simple derivation of GK based on the ordering of \secref{sec_ordering}
is given in \cite{Howes_etal}, along with a detailed GK treatment 
of the linear collisionless damping at $\kperp\rho_i\sim1$. 
{\em The GK is a valid approximation at all 
scales that are of interest in the context of low-frequency 
astrophysical turbulence, down to the electron gyroscale and below.} 
This broad range of validity and the long experience of GK 
simulations developed in fusion research make GK an ideal tool 
for numerical modeling of astrophysical turbulence\footnote{A programme 
of such numerical studies, using the {\tt GS2} code 
[http://gs2.sourceforge.net/], is currently 
underway (supported by the US DOE Center for Multiscale Plasma Dynamics).} 
and a good starting point for analytical theory. 

The RMHD and KRMHD equations (\secsref{sec_AW}{sec_nB}) 
can be derived from GK by means of two subsidiary expansions: first 
in $(m_e/m_i)^{1/2}$, then in $\kperp\rho_i\ll1$.\footnote{This means that 
the $\kpar/\kperp$, $(m_e/m_i)^{1/2}$ and $\kperp\rho_i$ expansions commute: 
KRMHD can be arrived at by either of the two routes: 
full kinetics $\to$ $\kperp\rho_i$ expansion $\to$ KMHD 
$\to$ $\kpar/\kperp$ expansion $\to$ $(m_e/m_i)^{1/2}$ expansion $\to$ 
isothermal electrons $\to$ KRMHD [this paper]
or 
full kinetics $\to$ $\kpar/\kperp$ expansion $\to$ GK \cite{Howes_etal} $\to$ 
$(m_e/m_i)^{1/2}$ expansion $\to$ isothermal electrons 
$\to$ $\kperp\rho_i$ expansion $\to$ KRMHD \cite{SCDHHQ_gk2}.} 
This and various other limits of the GK description 
of turbulence in weakly collisional astrophysical plasmas
are worked out in \cite{SCDHHQ_gk2}. While the $\kperp\rho_i\sim1$ 
regime requires solving the ion GK equation (electrons remain isothermal \cite{SCDHHQ_gk2}), 
the KAW turbulence at $\rho_i^{-1}\ll\kperp\ll\rho_e^{-1}$ is described 
by another well known fluid model, the Electron MHD (EMHD) \cite{Kingsep_Chukbar_Yankov}. 

\subsection{Kinetic Alfv\'en waves: Electron Reduced MHD} 
\label{sec_ERMHD}

When $\kperp\rho_i\gg1$, the ions are unmagnetized and have a
Boltzmann distribution: $f_i = \fMi(v)\exp\lt(-e\phi/T_0\rt)$, 
where $\phi$ is the scalar potential. 
The electrons are magnetized ($\kperp\rho_e\ll1$) 
and can be shown to be isothermal in 
essentially the same way as in \secref{sec_nB}, where  
\eqref{eq_dfe} is still valid to lowest order in $(m_e/m_i)^{1/2}$. 
Then 
\bea
\label{nBphi}
{\dne\over\ne} = {\dni\over\ni} = - {e\phi\over T_0} 
= -{1\over\beta_i}{\dBpar\over B_0}. 
\eea
The last equality follows from the (perpendicular) 
pressure balance similarly to the way it was done in \secref{sec_nB}, 
using $\dpperpi=\dpperpe=\dne T_0$ 
(the ordering of \secref{sec_ordering}, which eliminates the 
fast waves, continues to be valid). 
The EMHD equations now follow from the density and parallel velocity 
moments of the electron kinetic equation, which is similar in form 
to \eqref{eq_dfi}. The rigourous GK derivation is given in \cite{SCDHHQ_gk2}. 
Here, we adopt a more conventional approach by noting that 
in the limit $\kperp\rho_e\ll1$, the magnetic 
field is frozen into the electron fluid and satisfies \eqref{KMHD_B} 
with $\vu$ replaced by the electron flow velocity $\vu_e$ \cite{Kingsep_Chukbar_Yankov}:
\bea
\label{ind_eq_els}
{\dd\vB\over\dd t} = \vdel\times\lt(\vu_e\times\vB\rt) = 
-\vu_e\cdot\vdel\vB + \vB\cdot\vdel\vu_e - \vB\vdel\cdot\vu_e.
\eea
We set $\dvB/B_0=(1/v_A)\vz\times\vdperp\Psi + \vz\dBpar/B_0$ and  
expand \eqref{ind_eq_els} using the ordering of \secref{sec_ordering}. 
To lowest order in the $\kperp\rho_i\gg1$ expansion, 
$\vu_e$ can be found by taking the ions to be 
immobile and using Amp\`ere's law: 
$\vu_e = \vu_i - \vj/e\ne = - (c/4\pi e\ne)\vdperp\times\dvB$. 
In the last term in \eqref{ind_eq_els}, the next-order compressible 
part of $\vu_e$ is calculated via the electron continuity equation: 
$\vdel\cdot\vu_e = - (\dd/\dd t + \vu_e\cdot\vdperp)\dne/\ne$. 
Finally, using \eqref{nBphi} and denoting $\Phi=c\phi/B_0 = (cT_0/eB_0\beta_i)\dBpar/B_0$ 
[see \eqref{nBphi}], we 
find that the parallel and perpendicular components of \eqref{ind_eq_els} 
take the following form
\bea
\label{EMHD_Phi}
{\dd\Phi\over\dd t} = {v_A\over 2(1+\beta_i)}\,\Dpar\lt(\rho_i^2\dperp^2\Psi\rt),\\
\label{EMHD_Psi}
{\dd\Psi\over\dd t} = 2v_A\Dpar\Phi,
\eea
where $\Dpar$ is defined in \eqref{dd_def}. 
We shall refer to this system as {\em Electron Reduced MHD (ERMHD)} 
--- the anisotropic limit of EMHD.\footnote{
\Eqref{ind_eq_els} with $\vu_e = - (c/4\pi e\ne)\vdperp\times\dvB$ 
and $\vdel\cdot\vu_e=0$ (the incompressible limit valid if $\beta_i\gg1$) is what is 
normally understood by EMHD. In \eqsdash{EMHD_Phi}{EMHD_Psi}, $\beta_i$ is 
arbitrary, i.e., the electron fluid is not assumed to be exactly 
incompressible.}

ERMHD describes the cascade of {\em kinetic Alfv\'en waves (KAW)}, whose linear 
dispersion relation is $\omega=\pm\kpar v_A\kperp\rho_i/\sqrt{1+\beta_i}$ with eigenfunctions 
$\Phi\mp\kperp\rho_i\Psi/2\sqrt{1+\beta_i}$. To understand the nonlinear 
cascade, one may follow the spirit of GS theory, assuming anisotropy 
($\kpar\ll\kperp$) and strong interactions \cite{Biskamp_etal_EMHD2,Cho_Lazarian_EMHD}. 
This argument, reviewed at the end of \apref{ap_GS}, leads to a 
$\kperp^{-7/3}$ spectrum of magnetic fluctuations 
(note that for KAW-like fluctuations, 
$\dBpar/B_0\sim\Phi/\rho_i v_A \sim\kperp\Psi/v_A\sim\dBperp/B_0$) 
and to the relation $\kpar\sim\kperp^{1/3}$, quantifying 
the anisotropy. Both of these scalings have been confirmed by numerical 
simulations of EMHD \cite{Biskamp_etal_EMHD2,Cho_Lazarian_EMHD}. 
Note that the electric-field fluctuations 
in this regime should have a $\kperp^{-1/3}$ spectrum because 
$\dE\sim\kperp\phi\sim\kperp\rho_i(v_A/c)\dB$. Measurements of the spectra of $\dvB$ 
and $\dvE$ in the SW appear to corroborate these arguments \cite{Bale_etal} 
(see \figref{fig_bale}). 

The anisotropic KAW cascade is terminated at $\kperp\rho_e\sim1$ by the 
electron collisionless damping. The proper description of this process 
is again gyrokinetic \cite{Howes_etal}.

\section{Interstellar medium and the two regimes of MHD turbulence}
\label{sec_ISM}

\subsection{Weakly collisional limit}
\label{sec_weak_coll}

The anisotropic MHD turbulence in extrasolar plasmas is largely 
similar to the turbulence in the SW. The best studied of these plasmas 
is the interstellar medium (ISM), 
a hot low-density plasma ($\ne\sim 1~\mathrm{cm}^{-3}$, $T_0\sim10^4$~K 
for the Warm ISM phase) that makes up most of our and other galaxies' diffuse luminous 
matter. Turbulence in the ISM is stirred by colliding shock waves caused by 
supernova explosions, with the estimated injection scale 
$\lf\sim100$~pc~$\sim10^{20}$~cm \cite{Norman_Ferrara}. 
One important difference with the SW is that in the ISM, 
$\mfp\sim10^{12}$~cm is substantially smaller than $\lf$, although it is still 
larger than $\rho_i\sim10^9$~cm. Thus, collisions have to be allowed for. 
This can be done by keeping a collision integral in the GK 
equations and ordering the ion-ion collision rate to be comparable to 
the fluctuation frequency, $\nui\sim\omega$ \cite{Howes_etal,SCDHHQ_gk2}. 
Both the RMHD equations \exsdash{RMHD_Phi}{RMHD_Psi} 
above the ion gyroscale and the ERMHD equations \exsdash{EMHD_Phi}{EMHD_Psi} 
below it can then still be derived rigourously \cite{SCDHHQ_gk2}. 
Collisions do not appear in these 
equations: in \eqref{RMHD_Phi}, this is because in the 
$\kperp\rho_i\ll1$ limit the collisional transport is parallel 
to the field lines \cite{Braginskii}; in \eqref{RMHD_Psi}, 
the collision terms, which give rise to Ohmic resistivity, are ordered 
out via the subsidiary expansion in $(m_e/m_i)^{1/2}$; the latter is 
also true for the ERMHD equations \exsdash{EMHD_Phi}{EMHD_Psi}. 
The kinetic part of the KRMHD system, \eqsdash{eq_dfi}{KRMHD_B}, 
remains intact except that the ion-ion collision integral appears 
in \eqref{eq_dfi} \cite{SCDHHQ_gk2}. 
Thus modified, {\em the KRMHD constitutes a description of anisotropic 
plasma turbulence valid both in the collisional and collisionless 
regime.}\footnote{Strictly speaking, this is only true for $\kpar\mfp\gg(m_e/m_i)^{1/2}$. 
At longer parallel scales, the electrons are adiabatic, rather than 
isothermal, $\dTe=\dTi$, 
and the standard fluid MHD theory applies. With the ordering 
of \secref{sec_ordering}, the equations for the passive part of the 
turbulence are the same as \eqsdash{coll_dBpar}{coll_dTi}, except 
now $\dne/\ne = -\dTi/\Ti - (1/\beta_i)\dBpar/B_0$ \cite{SC_mhdbook,SCDHHQ_gk2} 
and the transport terms are more involved \cite{Braginskii}.} 
To lowest order in $\kperp\rho_i$, the collision integral has 
no spatial derivatives, so \eqref{eq_dfi} is still linear in the Lagrangian 
frame of the Alfv\'en waves and the discussion 
of the cascades of $\dne$ and $\dBpar$ given in \secref{sec_cascades} 
continues to apply. The only difference is that there is also 
collisional damping of these fluctuations, which, like the collisionless 
damping, depends solely on the variation of $\dne$ and $\dBpar$ along 
the perturbed magnetic-field lines. Indeed, in the collisional 
limit $\kparo\mfp\ll1$, \eqsdash{eq_dfi}{KRMHD_B} reduce to a set 
of fluid equations via the standard Chapman--Enskog expansion 
procedure \cite{SCDHHQ_gk2}:
\bea
\label{coll_dBpar}
{\rmd\over\rmd t}{\dBpar\over B_0} = \Dpar\upar + {\rmd\over\rmd t}{\dne\over\ne},\\
\label{coll_upar}
{\rmd\over\rmd t}\,\upar = v_A^2\Dpar{\dBpar\over B_0} + \nupar\Dpar\lt(\Dpar\upar\rt),\\
\label{coll_dTi}
{\rmd\over\rmd t}{\dTi\over\Ti} = {2\over3}{\rmd\over\rmd t}{\dne\over\ne} 
+ \kappar\Dpar\lt(\Dpar{\dTi\over\Ti}\rt),
\eea
where $\nupar\sim\kappar\sim\vthi\mfp$ are the parallel 
viscosity and thermal diffusivity.
The ion temperature is related to $\dne$ and $\dBpar$ 
via pressure balance, which is written in the form 
$\dne/\ne = -(1/2)\dTi/\Ti - (1/\beta_i)\dBpar/B_0$. 
\Eqsdash{coll_dBpar}{coll_dTi} describe passive cascades of 
slow waves ($\upar$ and $\dBpar$) and entropy fluctuations ($\dne$ and $\dBpar$) 
mixed by Alfv\'en waves \cite{Lithwick_Goldreich} 
via the nonlinearities contained in $\rmd/\rmd t$ and $\Dpar$ and damped by 
anisotropic diffusion, which occurs purely along the 
perturbed magnetic-field lines. 

The observational evidence is less exhaustive for the ISM than for the SW. 
The magnetic fluctuation spectra, inferred from the structure functions 
of the Faraday rotation measure, appear to be consistent with 
the $\kperp^{-5/3}$ scaling \cite{Minter_Spangler}, 
although the accuracy of the measurements is not high. 
The electron-density fluctuations, measured by a variety of methods, 
are anisotropic and also seem to have a Kolmogorov scaling across the 
entire range from $\lf\sim10^{20}$~cm to $\rho_i\sim10^9$~cm 
--- this is sometimes called ``The Great Power Law in the Sky'' 
\cite{Armstrong_Rickett_Spangler,Lazio_etal_review}.\POLAL{\footnote{There 
is, however, some evidence of a $\kperp^{-3/2}$ spectrum 
as well \cite{Smirnova_Gwinn_Shishov} --- see 
discussion of MHD turbulence scalings and the polarization-alignment 
theory \cite{Boldyrev} in \apref{ap_GS}.}} 
Note that while the density-fluctuation spectrum appears to 
extend to the ion gyroscale, the scale separation between 
$\mfp$ and $\rho_i$ is not sufficient in the ISM (unlike in the SW) 
to distinguish this from a cutoff at $\kpar\mfp\sim1$ (see \secref{sec_cascades}), 
which, using the GS relation $\kpar\sim\kperp^{2/3}L^{-1/3}$, would 
imply the perpendicular cutoff scale $\sim10^8$~cm \cite{Lithwick_Goldreich}. 

\subsection{Inhomogeneously turbulent ISM: spiral arms vs.\ interarm regions} 
\label{sec_arms}

It is, in fact, simplistic to view the ISM as a homogeneous plasma. 
The ISM is a spatially inhomogeneous environment consisting 
of several phases (of which Warm ISM is one) that have different temperatures, 
densities and degrees of ionization \cite{Ferriere_review} 
(and, therefore, different degrees of importance the neutral particles 
and the associated ambipolar damping effects have \cite{Lithwick_Goldreich}). 
While the role of the molecular properties of the multiphase ISM 
is left outside the scope of this paper, we would like to discuss briefly 
another aspect of the ISM's spatial inhomogeneity: the fact that it is 
inhomogeneously turbulent. One of the most prominent spatial features of 
our and many other galaxies is the spiral arms. 
They are denser than the interarm regions (interarms) by a factor 
of a few \cite{Roberts_Hausman} and observed to support stronger turbulence 
\cite{Rohlfs_Kreitschmann}, which is not surprising as the concentration 
of supernovae is higher. Observations of magnetic fields in external 
galaxies show that the spatially regular (mean) fields are stronger in 
the interarms, while in the arms, the stochastic fields dominate \cite{Beck_structure}. 
A recent study of the rotation-measure structure fuctions in 
our Galaxy \cite{Haverkorn_etal_arms} revealed 
that in the interarms, the magnetic energy is 
large-scale dominated and the structure functions are 
consistent with Kolmogorov-like negative spectral slopes, whereas 
in the arms, the structure functions are flat down to the resolution 
limit, meaning that the magnetic energy resides at much smaller scales 
than in the interarms. 
With these results in mind, let us recall that there exist 
two asymptotic regimes of MHD turbulence, depending 
on the relative magnitude of the mean and fluctuating fields, 
$\dBrms/B_0$: 

{\em I.~Anisotropic Alfv\'enic turbulence.} This is the type of turbulence 
discussed so far in this paper. It requires that a strong mean field 
$\vB_0$ is present. The turbulent fluctuations are much smaller than 
the mean field: $\dBrms\ll B_0$, $\urms\ll v_A$ 
(see \secref{sec_ordering}). The fluctuations are Alfv\'enic 
and have a Kolmogorov spectrum, 
with velocity and magnetic fields in scale-by-scale equipartition 
(see \apref{ap_GS}). 

{\em II.~Isotropic MHD turbulence.} In this case, no mean field is present, 
i.e., $B_0\ll\dBrms$. The dynamically strong stochastic 
magnetic field is a result of saturation of the {\em small-scale dynamo} 
--- amplification of magnetic field due to random stretching 
by the turbulent motions. Both the small-scale dynamo 
and its saturation are reviewed in \cite{SC_mhdbook}. While the definitive 
theory of the saturated state remains to be discovered, both physical 
arguments and numerical evidence \cite{SCTMM_stokes,YRS_exact} suggest 
that magnetic field is organized in folded flux sheets/ribbons. 
The length of these folds is comparable to the 
stirring scale, while 
the scale of the field-direction reversals transverse to the fold 
is determined by the dissipation physics: in MHD 
with Laplacian viscosity and resistivity operators, 
it is the resistive scale.\footnote{In weakly collisional astrophysical plasmas, 
such a description is not applicable 
and the field reversal scale is most probably determined by more complicated 
and as yet poorly understood plasma dissipation processes; below this scale, 
an Alfv\'enic turbulence of the kind discussed in \secref{sec_SW} may 
exist \cite{SC_dpp05}.} 
The structure functions of such magnetic fields are flat \cite{YRS_exact}, 
with magnetic energy dominantly at the reversal scale. 
While Alfv\'en waves 
propagating along the folds may exist \cite{SCTMM_stokes,SC_mhdbook}, 
the presence of small-scale direction reversals means that there is no 
scale-by-scale equipartition between velocity and magnetic fields. 

It is tempting to explain the difference between the magnetic-field 
structure in the arms and interarms by classifying the MHD turbulence in 
the interarms as anisotropic (I) and in the arms as isotropic 
(II). The observational evidence cited above lends qualitative 
support to this idea and so do numerical simulations of an inhomogeneously 
turbulent MHD fluid \cite{ICS_arms}. The turbulence in the arms should be 
closer to the isotropic variety and in the interarms to the anisotropic one 
for a number of conspiring reasons: 
(i) $\urms$ is larger in the arms, so $\dBrms/B_0 \sim \urms/v_A$ should be larger; 
(ii) the presence of the spiral mean field in galaxies 
is usually attributed to some form of mean-field 
dynamo \cite{Krause_Raedler}
and it is possible to argue plausibly that this mechanism 
produces stronger mean fields in the interarms than in the arms \cite{Shukurov_arms}; 
(iii) the mean field should be pushed out of the more turbulent region (arms) into the 
less turbulent one (interarms) by the diamagnetic effect of turbulence 
\cite{Zeldovich,Krause_Raedler,ICS_arms}. 

Finally, we mention another class of weakly collisional astrophysical plasmas 
where isotropic MHD turbulence is believed to exist: the intracluster 
medium (ICM) of the galaxy clusters. 
The turbulence in these intergalactic plasmas, 
which constitute the majority of the luminous matter 
in the Universe, has, in recent years, been increasingly accessible to 
observational astronomy \cite{Schuecker_etal,Vogt_Ensslin2}. 
For further information, 
the reader is referred to our review \cite{SC_dpp05}. 


\ack We gratefully acknowledge continued interactions with 
T.~En{\ss}lin, G.~Hammett, G.~Howes, A.~Iskakov, N.~Kleeorin, R.~Kulsrud, J.~McWilliams, 
J.~Mestel, E.~Quataert, F.~Rincon, I.~Rogachevskii, T.~Tatsuno, A.~Waelkens and T.~Yousef, 
who are involved in ongoing collaborations with us on the topics reviewed 
in this paper. We also thank S.~Chapman, M.~Haverkorn, B.~Hnat, T.~Horbury and A.~Shukurov  
for helpful discussions of observational evidence, 
and S.~Bale for permission to use figure 3 of \cite{Bale_etal}.
A.A.S.\ was supported by a PPARC Advanced Fellowship and 
by King's College, Cambridge. 
This work has benefited from support by the US DOE Center for 
Multiscale Plasma Dynamics. 

\appendix

\section{Scaling theories of Alfv\'en-wave turbulence: a brief review}
\label{ap_GS}

\subsection*{Goldreich--Sridhar turbulence}

Here we outline the key steps in the GS theory of anisotropic 
MHD turbulence \cite{GS95}. A more leisurely historical review, which explains 
how the GS argument is related to the earlier (isotropic) 
theory of Iroshnikov \cite{Iroshnikov} and Kraichnan \cite{Kraichnan} 
and to the weak-turbulence treatment \cite{GS97,Galtier_etal}, 
can be found in \cite{SC_mhdbook} (see also \cite{Galtier_Pouquet_Mangeney} 
and \cite{Lithwick_Goldreich_Sridhar}). 

As in the Kolmogorov--Obukhov theory of turbulence, it is 
assumed that the cascade of energy is local in scale space and 
the flux of energy through scale $\lambda$ in the inertial range 
is scale-independent: 
\bea
{\ul^2\over\taul}\sim\varepsilon = \const,
\label{const_flux}
\eea
where $\varepsilon$ is the Kolmogorov flux, 
the subscript $\lambda$ indicates fluctuations associated with 
the perpendicular scale $\lambda$, 
and $\taul$ is the cascade time. 
It is now assumed that the turbulence is 
strong, i.e., that the Alfv\'enic linear propagation terms are 
comparable to the nonlinear terms:
\bea
v_A\dpar \sim \vuperp\cdot\vdperp 
\quad\Leftrightarrow\quad
{v_A\over\lparl}\sim{\ul\over\lambda},
\label{lin_nlin}
\eea
This is the critical-balance conjecture, applied scale by scale. 
It is further assumed that 
the cascade time is the same as the Alfv\'en time: 
$\taul\sim {\lparl/v_A}$. Together with \eqsdash{const_flux}{lin_nlin}, 
this immediately implies 
\bea
\ul\sim\lt({\varepsilon\lparl\over v_A}\rt)^{1/2}\sim \lt(\varepsilon\lambda\rt)^{1/3},\quad
\lparl\sim \lt({v_A^3\over\varepsilon}\rt)^{1/3}\lambda^{2/3}. 
\label{GS_scaling}
\eea
The first of these scaling relations is equivalent to 
a $\kperp^{-5/3}$ spectrum of kinetic energy,\footnote{In terms 
of parallel wavenumbers, \eqref{GS_scaling} means that 
the spectrum scales as $\kpar^{-2}$. Remarkably, recent SW data 
analysis confirms this power law \cite{Horbury_etal_aniso}.} 
the second quantifies the anisotropy by 
establishing the relation between the perpendicular 
and parallel scales. The fluctuations are Alfv\'enic, so 
$\dBpl\sim\ul\sqrt{4\pi\rho_0}$.  

\POLAL{
\subsection*{Polarization alignment}

While the GS theory  
has acquired the status of the accepted view, 
the failure of the numerical simulations \cite{Maron_Goldreich,Mueller_Biskamp_Grappin}
to reproduce the $\kperp^{-5/3}$ spectrum has remained a worrying 
puzzle. The numerical spectra are closer 
to $\kperp^{-3/2}$, but cannot be explained 
by the Iroshnikov--Kraichnan theory \cite{Iroshnikov,Kraichnan} 
because the fluctuations are definitely anisotropic. 
Recently, Boldyrev \cite{Boldyrev} proposed a scaling argument 
that allows an anisotropic Alfv\'enic turbulence with a $\kperp^{-3/2}$ spectrum. 
It is based on the conjecture that 
$\vuperp$ and $\dvBperp$ align at small scales, 
an idea that has had some numerical support 
\cite{Maron_Goldreich,Beresnyak_Lazarian,Mason_Cattaneo_Boldyrev}. 
The alignment weakens nonlinear interactions and alters the scalings. 

The fluctuations are assumed to be three-dimensionally 
anisotropic: the three characteristic scales are 
the parallel scale $\lpar$ along $\vB_0$, 
the displacement $\xip$ of the fluid element 
perpendicular to $\vB_0$ and parallel to $\dvBperp$, 
and the scale $\lambda$ of the variation of 
$\vuperp$ and $\dvBperp$ 
perpendicular both to $\vB_0$ and to $\dvBperp$. 
The nonlinear terms in \eqsdash{RMHD_Phi}{RMHD_Psi} are
\bea
\dvBperp\cdot\vdperp\sim {\dBpl\over\xil},\quad
\vuperp\cdot\vdperp \sim {\ul\thl\over\lambda}\sim {\ul\over\xil},
\label{nlin_estimate}
\eea
where $\thl$ is the angle between 
$\vuperp$ and $\dvBperp$, assumed to be small, and $\vdperp\cdot\vuperp=0$ 
has been used to estimate $\thl\sim\lambda/\xil$, 
which is, indeed, small if $\xil\gg\lambda$. 

Further development is the same as in the Kolmogorov/GS argument reviewed above, 
except that in \eqref{lin_nlin} and, consequently, in \eqref{GS_scaling}, 
$\lambda$ must be replaced by $\xil$. 
An additional assumption is now needed to determine $\xil$. 
Boldyrev conjectures that $\vuperp$ and $\dvBperp$ 
will align to the maximum possible extent. This 
is achieved if the angle $\thl$ between 
them is comparable to the characteristic angular wonder 
of $\dvBperp$: 
\bea
\thl\sim{\lambda\over\xil}\sim{\xil\over\lparl}
\quad\Rightarrow\quad
\xil\sim\lt(\lambda\lparl\rt)^{1/2}.
\label{pol_al}
\eea
Combining \eqref{GS_scaling} (where $\lambda$ is replaced by $\xil$) 
and \eqref{pol_al}, one gets  
\bea
\label{pol_al_spec}
\ul\sim\lt({\varepsilon\lparl\over v_A}\rt)^{1/2}
\sim\lt(\varepsilon\xil\rt)^{1/3} 
\sim\lt(\varepsilon v_A \lambda\rt)^{1/4},\\
\xil\sim \lt({v_A^3\over\varepsilon}\rt)^{1/4}\lambda^{3/4},\quad
\lparl\sim \lt({v_A^3\over\varepsilon}\rt)^{1/3}\xil^{2/3}
\sim \lt({v_A^3\over\varepsilon}\rt)^{1/2}\lambda^{1/2}. 
\label{pol_al_scales}
\eea
The scaling relation \exref{pol_al_spec} is equivalent to 
a $\kperp^{-3/2}$ spectrum of kinetic energy. 

The status of Boldyrev's theory vis-{\`a}-vis 
real MHD turbulence is uncertain. 
Observationally, only in the SW does one
measure the spectra with sufficient accuracy to state that they 
are consistent with $\kperp^{-5/3}$ but {\em not} with $\kperp^{-3/2}$ 
\cite{Leamon_etal,Bale_etal,Horbury_etal_review}. 
From numerical simulations, 
it appears that the condition for the $\kperp^{-3/2}$ spectra 
\cite{Maron_Goldreich,Mueller_Biskamp_Grappin} and 
the alignment scaling $\thl\sim\lt({\varepsilon/v_A^3}\rt)^{1/4}\lambda^{1/4}$ 
\cite{Mason_Cattaneo_Boldyrev} 
to emerge is that the mean field is strong 
(a few times $\dBrms$),\footnote{Note however, that \cite{Maron_Goldreich} 
had $B_0\sim100\dBrms$, reported a $\kperp^{-3/2}$ spectrum, 
but also found that the anisotropy fit the GS scaling 
$\lparl\sim\lambda^{2/3}$, not $\lparl\sim\lambda^{1/2}$
that appears in \eqref{pol_al_scales}.} 
whereas in the SW, $B_0\sim\dBrms$. 
It is not, however, clear why that should matter asymptotically, 
because $\dBpl/B_0$ is arbitrarily small for sufficiently small~$\lambda$. 
}

\subsection*{Scaling of passive scalar fields} 

The scaling of the passively mixed scalar fields, e.g., 
density fluctuations $\dne$, 
is slaved to the scaling of the Alfv\'enic fluctuations. 
Again as in Kolmogorov--Obukhov theory, one assumes 
a local-in-scale-space cascade of scalar variance 
and a constant flux $\epsn$ of this variance. 
Then, analogously to \eqref{const_flux}, 
$\dnl^2/\taul\sim\epsn$. The cascade time is 
$\taul^{-1}\sim\vuperp\cdot\vdperp\sim v_A/\lpar\sim\varepsilon/\ul^2$. 
This gives 
\bea
\dnl\sim\lt({\epsn\over\varepsilon}\rt)^{1/2}\ul, 
\eea
so the scalar fluctuations have the same scaling as 
the turbulence that mixes them. 

\subsection*{Kinetic-Alfv\'en-wave turbulence}

The scaling laws for the KAW turbulence are again obtained following 
the Kolmogorov--Obukhov/GS line of reasoning 
\cite{Biskamp_etal_EMHD2,Cho_Lazarian_EMHD}. 
Locality of interactions and constancy of the energy flux imply, 
analogously to \eqref{const_flux},
\bea
\lt({\dBl\over B_0}\rt)^2{v_A^2\over\taul} \sim \epsB = \const.
\label{const_flux_KAW}
\eea
If the turbulence is strong, then, analogously to \eqref{lin_nlin},
\bea
{\dd\over\dd z} \sim {\dvBperp\over B_0}\cdot\vdperp
\quad\Leftrightarrow\quad {\dBl\over B_0}\sim{\lambda\over\lparl}.
\label{lin_nlin_KAW}
\eea
Assuming that the cascade time is comparable to the inverse KAW frequency, 
$\taul\sim {\lparl\lambda/v_A\rho_i}$,  
and combining this with \eqsdash{const_flux_KAW}{lin_nlin_KAW}, 
we get
\bea
{\dBl\over B_0} 
\sim \lt({\epsB\over v_A^3\rho_i}\rt)^{1/3}\lambda^{2/3},\quad
\lparl \sim \lt({v_A^3\over\epsB}\rt)^{1/3}\rho_i^{1/3}\lambda^{1/3}.
\eea
The first of these scaling relations is equivalent to a $\kperp^{-7/3}$ 
spectrum of magnetic energy, the second quantifies the anisotropy. 
Note that for KAW-like fluctuations, $\dBparl\sim\dBpl\sim\dBl$,  
$\dE_\lambda\sim(v_A\rho_i/c)\dBl/\lambda$ and $\dnl/\ne\sim\dBl/B_0$ 
(see \secref{sec_ERMHD}).

\end{document}